# Localization microscopy: a review of the progress in methods and applications

Jack W Shepherd[a,b], Mark C Leake[a,b,*]

[a]Department of Physics, University of York, York, YO10 5DD

[b]Department of Biology, University of York, York, YO10 5DD

*Email: mark.leake@york.ac.uk

**Abstract**

Here, we report analysis and summary of research in the field of localization microscopy for optical imaging. We introduce the basic elements of super-resolved localization microscopy methods for PALM and STORM, commonly used both *in vivo* and *in vitro*, discussing the core essentials of background theory, instrumentation and computational algorithms. We discuss the resolution limit of light microscopy and the mathematical framework for localizing fluorescent dyes in space beyond this limit, including the precision obtainable as a function of the amount of light emitted from a dye, and how it leads to a fundamental compromise between spatial and temporal precision. The properties of a "good dye" are outlined, as are the features of PALM and STORM super-resolution microscopy and adaptations that may need to be made to experimental protocols to perform localization determination. We analyse briefly some of the methods of modern super-resolved optical imaging that work through reshaping point spread functions and how they utilize aspects of localization microscopy, such as stimulated depletion (STED) methods and MINFLUX, and summarize modern methods that push localization into 3D using non-Gaussian point spread functions. We report on current methods for analyzing localization data including determination of 2D and 3D diffusion constants, molecular stoichiometries, and performing cluster analysis with cutting-edge techniques, and finally discuss how these techniques may be used to enable important insight into a range of biological processes.

**Keywords**

## 1. The optical resolution limit

Antony van Leeuwenhoek (1632-1723) was the pioneer of the light microscope [1]. Using glass beads with radii of curvature as small as 0.75 mm taken from blown or drawn glass, he managed to construct the seminal optical microscope, with a magnification of 275x and spatial resolution only slightly above one micron. Granted, the microscope had to be in effect jammed into the user's eye and the sample held fractions of an inch from the lens, but it is remarkable that *ca.* 350 years ago a simple light microscope existed which had a spatial resolution only 3-4 times lower than a the so-called "diffraction-limited barrier", subsequently known as the optical resolution limit.

Why is it, then, that such a "diffraction limit" exists? The answer emerged ~150 years after van Leeuwenhoek's death from the theoretical deliberations of the German polymath Ernst Abbe: he argued that, when we image an object due to its scattering of light at a large distance from it ("large" being greater than several wavelengths of light), optical diffraction reduces sharpness in direct proportion to the wavelength of light used and in inverse proportion to the largest possible cone of light that can be accepted by the objective lens used (this quantity is characterized by the numerical aperture, or "NA"). Expressed algebraically we find that the minimum resolvable distance using ordinary light microscopy assuming imaging through a rectangular aperture is

$$d = \frac{\lambda}{2\text{NA}}$$

A variant of this formula as we will see below includes a factor of 1.22 in front of the wavelength $\lambda$ parameter to account for circular apertures as occur in traditional light microscopy. Visible light has a wavelength approximately in the range 400-700 nm, and the best objective lenses commonly used in single objective lens research microscopes, at least those that avoid toxic organic solvent immersion oil, have an NA of around 1.5, implying that this Abbe limit as denoted above is somewhere between 100-250 nm, larger than many viruses but good enough to easily visualize bacteria, mammalian cells, and archaea as well as several subcellular features such mitochondria, chloroplast, nuclei, endosomes and vacuoles. However, this spatial resolution is clearly not sufficient to observe the activity of the cell on a single-molecule level whose length scale is 1-2 orders of magnitude smaller.

Fortunately, this apparent hard limit can be softened: if one images an object that has a known shape (or at least that has a shape that has a known functional form) then we may fit an approximate mathematical model to the image obtained from the microscope. A parameter of this fit will be the intensity centroid of the object – and this is the key feature of "localization microscopy". This centroid may be expressed to a sub-pixel resolution albeit with a suitable error related to parameters such as the number of photons sampled, the noise of the detector and the size of the detector pixels. In brightfield imaging this principle is commonly used to track beads attached to filamentous molecules for tethered particle motion experiments, for example; the first reported use of Gaussian fitting for localization microscopy was actually in 1988 by Jeff Gelles and co-authors, who found the intensity centroid of a plastic bead being rotated by a single kinesin motor to a precision of a few nanometers [2]. With the added binding specificity potential of fluorescence labelling and subsequent imaging, localization microscopy can go much further though.

## 2. Super-resolved localization in 2D

Fluorophores imaged onto the Cartesian plane of a 2D camera detector are manifest as a characteristic point spread function (PSF) known as the Airy disk, consisting of an intense central Gaussian-like zeroth order peak surrounded by higher order concentric rings of intensity, as shown in Figure 1a. Physically, the concentric rings arise due to Fraunhofer diffraction as the light propagates through a circular aperture. Mathematically, the intensity distribution due to this effect is given by the modulus of the Fourier transform of the aperture squared. The Rayleigh criterion (though note there are other less used resolution criteria that could be used, such as the Sparrow limit) specifies that the minimum separation of two resolvable Airy disks is when the intensity peak of one coincides with the first intensity minimum of the other, and for circular lenses we find

$$d = \frac{0.61\lambda}{\text{NA}}$$

For example, for two green fluorescent protein (GFP) fluorophores [3], a very common dye use in live cell localization microscopy, emitting at a peak wavelength of 507 nm under normal physiological conditions and a typical 1.49 NA objective lens this value of *d* is 208 nm and therefore to obtain spatial localization information for more than one molecule they must be separated by at least this distance, or alternatively emitting at different times such that each molecule can be analyzed separately.

| Algorithm Name | Description | Features | Pros | Cons | Notes |
|---|---|---|---|---|---|
| Haar wavelet kernel (HAWK) analysis [4] | Data pre-processing method | Decomposes high-density data to create longer, lower density dataset for further analysis | May not collapse structures ~200 nm apart into one structure; experiments easier as high density can be worked around | Depending on imaging, the localization precision may be comparable to lower resolution techniques e.g. structured illumination microscopy (SIM) [5] | Versatile: can be used with any frame-by-frame localization algorithm |
| DAOSTORM [6] | STORM-type analysis package | Fits multiple PSF model to pixel cluster to deal with overlapping fluorophores | Increases workable density from ~ 1 to ~7 molecules/μm$^2$ dependent on conditions | Requires long STORM-type acquisitions; no temporal information | Adapted from astronomy software |
| FALCON [7] | Localization for STORM/PALM-type data | Iteratively fits Taylor-series expanded PSFs to data to find best fit | Data treated as a continuum rather than on a grid | Low temporal resolution (~2.5 s/frame) | Localization error *ca.* 10 to 100 nm depending on imaging conditions |
| ThunderSTORM [8] | PALM/STORM localization | Toolbox of analysis algorithms user can choose from | Flexible | Many methods need careful selection of parameter values, though some can be set algorithmically | Free ImageJ plugin. Also functions as a data simulator for testing routines |
| Bayesian analysis of blinking and bleaching (3B) [9] | Factorial hidden Markov model which models whole system as a combination of dark and emitting fluorophores | Produces a probabilistic model of the underlying system; exact positions may have high error bars | Can analyse overlapping fluorophores, eliminates need for traditional user-defined parameters, no dependence on specialized hardware | Results require nuanced interpretation, Bayesian priors must be well known, resolution may vary along and perpendicular to a line of fluorophores | Spatial resolution *ca.* 50 nm; temporal ~s |
| ADEMSCode [10] | Matlab toolkit for single-molecule tracking and analysis described in this Section | Can analyze stoichiometries, diffusion coefficients, localization | Powerful, flexible | Human-selected parameters require careful treatment | Accesses high temporal resolution (~ms) as well as localization |
| Single-Molecule Analysis by Unsupervised Gibbs sampling (SMAUG) [11] | Bayesian approach analysing trajectories of single molecules | Analyzes trajectories to find underlying mobility states and probabilities of moving between states | Non-parametric, reduces bias | Reliant on the quality of input data, needs good priors | Post-processing – input data generated by other techniques |
| RainSTORM [12] | MATLAB image reconstruction code | Complete STORM workflow including data simulation | Simple, out-of-the box operation | Weaker with dense data, relatively inflexible, no dynamical information | |
| QuickPALM [13] | PALM analysis software | 3D reconstruction, drift correction | Complete PALM solution | Static reconstruction | Plugin for ImageJ [14] |

**Table 1: Comparison of some modern super-resolving localization microscopy analysis packages.**

If we meet these conditions, we will generate an image similar to that in Figure 1b. The diffraction-limited fluorescent "spot" (essentially the zeroth order peak of the Airy disk) is clearly visible spread over multiple pixels, though there is significant background noise, and taking line profiles shows that it is approximately Gaussian in both *x* and *y* Cartesian axes. One way to proceed with localization determination using a computational algorithm is the following, exemplified by software that we have developed called ADEMSCode [10], but with a plethora of similar algorithms used by others in this field (see Table 1), including probabilistic Bayesian approaches (e.g. 3B [9]) and pre-processing steps to reduce fluorophore density and improve the effectiveness of subsequent analysis [4]. Most of these are capable of analyzing fluorophore localizations, but to our knowledge our own package [10] is the only one which is capable of evaluating localizations, dynamical information such as diffusion coefficients, and utilizing photobleaching dynamics to estimate molecular stoichiometries. ADEMSCode proceeds with localization analysis in the following way: first, find the peak pixel intensity from the camera image, and draw a small bounding box around it (typically 17x17 pixels (i.e. one central pixel with a padding of 8 pixels on each side), where for us a pixel is equivalent to 50-60 nm at the sample). Within that square then draw a smaller circle (typically of 5 pixel radius) centered on the maximum of intensity which approximately contains the bulk information of the fluorophore's PSF. The pixels which are then within the bounding box but not within the circle may have their intensities averaged to generate a local background estimate. Each pixel then has the local background value subtracted from it to leave the intensity due only to the fluorophore under examination, and this corrected intensity may now be fitted. An optimization process then occurs involving iterative Gaussian masking to refine the center of the circle and ultimately calculate a sub-pixel precise estimate for the intensity centroid. A similar effect can be achieved by fitting a 2D Gaussian function plus uniform offset to intensity values that have not been background corrected, however, fit parameters often have to be heavily constrained in the low signal-to-noise regimes relevant to imaging single dim fluorescent protein molecules and due to the centroid output, iterative Gaussian masking is often more robust.

Although it has an analytic form, for historical reasons relating to past benefits to computational efficiency, the central peak of the Airy disk is commonly approximated as a 2D Gaussian that has equation

$$I(x,y) = I_0 e^{-\left(\frac{(x-x_0)^2}{2\sigma_x^2} + \frac{(y-y_0)^2}{2\sigma_y^2}\right)}$$

where the fittable parameters are $I_0$, the maximum brightness of the single fluorophore, $x_0$ and $y_0$, the co-ordinates of the center of the Gaussian, and $\sigma_x$ and $\sigma_y$ which are the Gaussian widths in $x$ and $y$ respectively (interested readers should read the work of Kim Mortensen and co-workers on the improvements that can be made using a more accurate formulation for the PSF function [15]). Using conventional fluorescence microscopy, assuming any potential polarization effects from the orientation of the dipole axis of the fluorophore are over a time scale that is shorter that the imaging time scale but typically 6-7 orders of magnitude, the Airy disk is radially symmetrical, and so $\sigma_x$ and $\sigma_y$ ought to be identical, and they may therefore be used as a sanity check that there is only one molecule under consideration – a chain of individually unresolvable fluorophores will have a far higher spread in $x$ or $y$. Similarly, the brightness of individual fluorophores in a dataset acquired under exactly the same imaging conditions is a Gaussian distribution about a mean value. After fitting the 2D Gaussian one may usefully plot the fitted $I_0$ values to check for outliers; indeed, with a well-characterized fluorophore and microscope one may be able to include these checks in the analysis code itself. When iterative masking is used to determine the intensity centroid an initial guess is made for the intensity centroid, and a 2D function is then convolved with the raw pixel intensity data in the vicinity of each fluorescent spot. These convolved intensity data then have a revised intensity centroid, and the process is iterated until convergence. The only limitation on the function used is that it is radially symmetric, and it has a local maximum at the center. In other words, a triangular function would suffice, if the purpose is solely to determine the intensity centroid. However, a Gaussian function has advantages in returning meaningful additional details such as the sigma width and the integrated area.

Having fit the 2D Gaussian, the fitting algorithm will usually report to the user not only the best-fit values but also either estimated errors on these fits or the matrix of covariances which may be trivially used to obtain the error bars. It is tempting to take the error on $x_0$ and $y_0$ optimized fitting values as the localization precision, but this reflects the fitting precision that only partially indicates the full error involved. In fact, the error on the centroid needs to be found by considering the full suite of errors involved when taking the experimental measurement. Principally, we must include the so-called "dark noise" in the camera, the fact that we cannot know which point in a pixel the photon has struck and therefore the PSF is pixelated, and the total number of photons that find their way from the fluorophore to the sensor. Mathematically, the formulation is given in reference [16] as

$$\delta = \sqrt{\left(\frac{s^2 + \frac{(lm)^2}{12}}{N}\right) + \left(\frac{4\sqrt{\pi}s^2 b^2}{lmN^2}\right)}$$

where $s$ is the fitted width of the Gaussian PSF and would usually be taken as a mean of $\sigma_x$ and $\sigma_y$, $N$ is the number of photons, $b$ is the camera dark noise, $l$ is the camera pixel edge length and $m$ is the magnification of the microscope. We can instantly see that a compromise must be reached during experiment. If we image rapidly we will have fewer photons to fit to and the spatial localization precision worsens. The approximate scaling is with the reciprocal of the square root of the number of photons sampled in the case of relatively low dark noise and small camera pixels. If we image for a long time we can see localization precisions as low as one nanometer at the cost of losing dynamical information. In practice we find that imaging on the order of millisecond exposures leads to lateral (i.e. In the 2D plane of the camera detector) localization precisions of 30-50 nm for relatively poor dyes such as fluorescent protein molecules, while imaging for a second or more on far bright organic dyes may give single nanometer precision [17]. This fundamental trade-off has led to two complementary but different forms of super-resolved localization microscopy – long time-course imaging on fixed cells for nm precision localization, and lower spatial resolution imaging that can access temporal information.

If we have obtained a time-series acquisition of a system with mobile fluorophores (either freely diffusing or attached to a translocating molecular machine, for example) we may wish to work out where and how quickly the fluorescent spots are moving, or if their mobility is Brownian or 'anomalous' or confined. The localization information may then be used to infer the underlying types of single-molecule mobility [18]. With localizations in hand this is relatively straightforward and is achieved by comparing successive frames ($n$ and $n + 1$) and accepting a that a spot in frame $n + 1$ is the same molecule as a spot in frame $n$ if the two spots are sufficiently close together, have sufficiently comparable intensities, and are of sufficiently comparable shape. The vector between the spots may then be taken and the process repeated for frames $n + 1$ and $n + 2$, iteratively building up a 2D track. This is a threshold-based method and should therefore be used with care, with the threshold determined by converging one physical parameter like diffusion coefficient with respect to the distance cut-off (a useful review of this effect is found in reference[19]). There should also be sufficiently few fluorophores such that a spot in frame $n + 1$ is not (or only rarely could be) accepted as being the same molecule as two or more spots in frame $n$ (or *vice versa*); should that occur the track will need to be terminated at that point. Deciding whether spots in two successive frames are the same molecule is clearly fraught with danger; modern methods with Bayesian analysis will be discussed later in our study here.

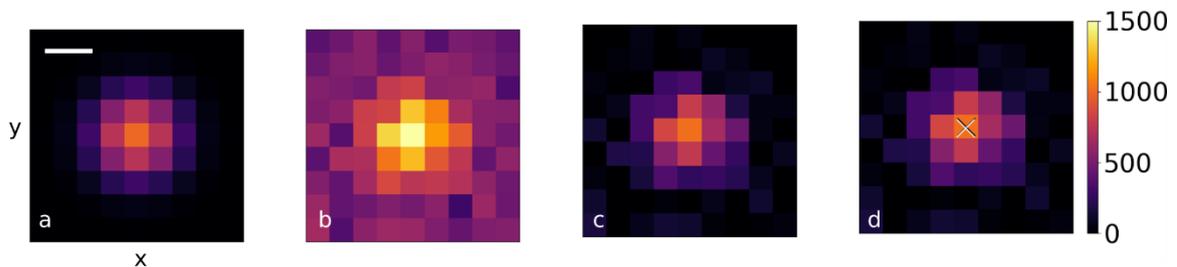

Figure 1. a) a Gaussian PSF; b) A Gaussian PSF as seen with background noise; c) the Gaussian PSF with noise after background correction; d) The results of fitting a 2D Gaussian to c. White cross is the center of the fit Gaussian and the black cross is the true center of the PSF. Bar: 100 nm.

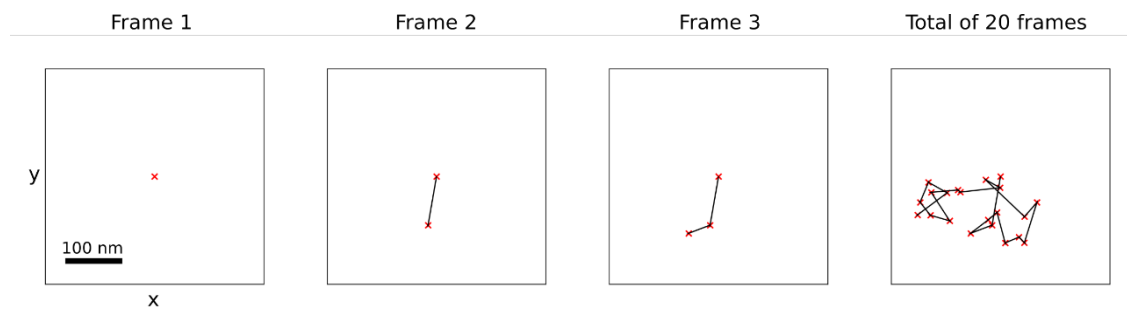

Figure 2. Combining localizations from multiple frames to build a 2D track.

### 3. STORM and PALM

Though STORM (STochastic Optical Reconstruction Microscopy) and PALM (PhotoActivated Light Microscopy) have differences in their methods, they work towards the same goal: to spatially and temporally separate the "on" states of the fluorophores used so they the PSFs can be fitted to as described above. By fitting a large population of fluorophores an image of the overall structure or distribution of the system of interest may be generated. Here we will briefly describe each technique and their relative pros and cons.

1. STORM

STORM (STochastic Optical Reconstruction Microscopy) [20] is a powerful technique which relies on the inherent ability of some fluorophores to switch between "on" (emitting) and "off" (non-emitting) states when irradiated by a high intensity laser. At the beginning of the image acquisition almost all fluorophores will be in the on state. However, as time goes on, the population gradually moves to a combination of fluorophores that have photobleached and are permanently non-emissive, and some that are in the photoblinking state in which they transition between on and off states. At some point, the ratio of these populations will reach the correct state such that individual fluorophores are visible separated from their neighbors, i.e. the mean nearest-neighbor distance between photoactive fluorophores is then greater than the optical resolution limit and so a distinct PSF image associated with each separate fluorophore molecule can be seen. A time series of frames must be acquired of the system in these conditions, and each fluorophore in each frame is localized as described above. The loci found may then be all be plotted in one frame, showing the base distribution of the fluorophores and thus the structure of the system. This is "optical reconstruction" – although in each image frame only a few individual foci may be visible, by combining several hundreds or thousands of these image frames enough fluorophores will be captured to give the overall picture in detail far beyond non super-resolution microscopy methods.

For this process to be feasible, several conditions must be met. First, the excitation laser power must be high enough to force the fluorophores into their photoblinking state. Though it is counterintuitive, a low power laser will enable the fluorophores to stay on longer but photobleach permanently afterwards, without any blinking. In general, laser excitation intensities at or above *ca.* 1800 W/cm$^2$ are effective depending on the dyes used. Secondly, the fluorophores of interest should be capable of photoblinking behavior, and when they do blink their single-molecule brightness must be above the background noise. Fluorophores which are suitable under these constraints will be discussed in Section 4.

2. PALM

PALM (PhotoActivated Light Microscopy) [21] takes a second approach to separating fluorophore emissions in space and time. While STORM relies on all fluorophores being excited at the same time but randomly blinking on and off, PALM randomly activates a random subset of the fluorophores in the system with one laser, and then excites them for imaging with a second laser. Activated fluorophores return to the initial state after they are imaged. Then repeat, and image a second set of fluorophores. Activation can mean either one of two processes. Either the fluorophore is initially dark and switches to a fluorescent state, or under illumination of the activation laser the fluorophore undergoes a color change, commonly from red to green. In either case the activating laser is usually ~long UV at around 400 nm wavelength, while the fluorescence excitation laser is in the ordinary visible range.

The constraints for PALM fluorophores are obvious. Although they do not need to photoblink, they must be capable of switching states in response to UV light exposure, and once again they must be bright enough in their emissive state to be well above the background noise level. As PALM images single molecules, the laser intensity must be relatively high as for STORM.

3. Pros and cons

STORM and PALM are powerful techniques that enable reconstruction of tagged systems, for example microtubules *in vivo* or the architecture of organelles in the cell. In this respect, the information offered by STORM and PALM is unrivalled by other techniques – more detailed information is difficult to find as the crowded cellular environment precludes whole-cell X-ray or neutron diffraction experiments. Imaging tagged substructures using traditional diffraction-limited optical microscopy is possible but gives less detailed information, and even mathematical post-processing techniques such as deconvolution (if they are suitable for the imaging conditions) give a lower resolution than super-resolution imaging itself.

However, these are not "magic bullet" techniques and have their own drawbacks. Principally, both are slow methods. To collect enough information to properly reconstruct the base fluorophore distribution hundreds to thousands of frames at least must be taken, meaning that total imaging times are seconds to minutes. Given that many biological processes occur over millisecond timescales or faster this obviously precludes capturing time-resolved information from these rapid dynamic processes. Further, if there is some biological process restructuring the cellular environment during imaging a false picture may be obtained. For this reason, biological samples are usually "fixed" i.e. rendered static and inert before imaging to ensure that the fluorophore distribution does not change during image acquisition. Photodamage is also of concern. As fluorophores photobleach, they produce free radicals which attack and damage the biological sample. Various imaging buffers exist which minimize this though these can induce lower photoblinking, so a tradeoff must be struck.

4. Techniques using modified point spread functions that use localization microscopy at some level

STORM and PALM are both powerful techniques in their own right but they are not the only way to generate data that can be processed with a super-resolution algorithm. In 2000, Stefan Hell (who went on to share the 2014 Nobel Prize in Chemistry with William E. Moerner and Eric Betzig for "the development of super-resolved fluorescence microscopy" [22]) published an account of a new super-resolution method based around stimulated emission of fluorophores, and known as STED (Stimulated Emission Depletion) microscopy [23], [24]. In brief, STED involves two lasers that are focused on the same position, one that excites the fluorophores while a donut-shaped beam around this has the effect of suppressing emission from fluorophores in this region, achieved via stimulated emission when an excited-state fluorophore interacts with a photon whose energy is identical to the difference between ground and excited states. This molecule returns to its ground state via stimulated emission before any spontaneous fluorescence emission has time to occur, so in effect the fluorescence is depleted in a ring around the first laser focus. By making the ring volume arbitrarily small the diameter of this un-depleted central region can be made smaller than the standard diffraction-limited PSF width, thus enabling super-resolved precision using standard localization fitting algorithms to pinpoint the centroid of this region, <30 nm being typical at video-rate sampling of a few tens of Hz.. Other related STED-like stimulated depletion approaches include Ground State Depletion (GSD) [25], saturated pattern excitation microscopy (SPEM) [26] and saturated structured illumination microscopy (SSIM) [27]. A similar result to STED using reversible photoswitching of fluorescent dyes but not reliant on stimulated emission depletion is known as RESOLFT (reversible saturable/switchable optical linear (fluorescence) transitions) microscopy [28] that also utilizes localization microscopy algorithms.

Similar to but going beyond STED approaches is a recently developed method also from Stefan Hell and colleagues known as MINFLUX (MINimal photon FLUXes) [29] which does not need a depletion laser. Here, the excitation beam is the donut and so a fluorophore at the center of the beam will not be excited. By scanning the beam and finding the intensity minimum, the position of a fluorophore can then be found with a nanoscale spatial precision small spatial precision with an exceptionally fast scan rate of up to *ca.* 10 kHz.

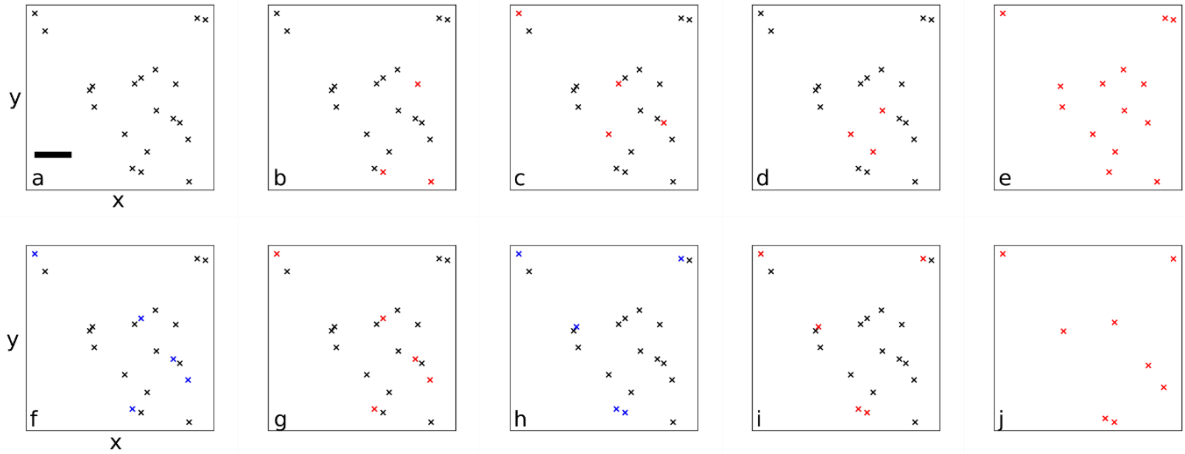

a. **Figure 3: Schematic of STORM and PALM localization. A)** the underlying fluorophore distribution. **b-d)** fluorophores are stochastically excited during STORM. **e)** reconstructing the original distribution from the emissions observed. **f-i)** In PALM, fluorophores are first activated with a UV laser (activated fluorophores in blue) and are then excited to fluoresce (red). **j)** The underlying distribution reconstructed. A given experiment time will produce fewer emission events in PALM and thus sample the underlying distribution slower than STORM. Bar: 200 nm.

4. **Choosing a fluorophore**

Selecting the correct fluorophore for the system of interest is clearly a prime concern. Summarized, fluorophores must:

- be bright enough for single molecules to be seen above background noise
- be photoactivatable or photoblink under the correct conditions
- not interfere with ordinary cell processes if imaging *in vivo*
- not unduly change the structure or function of an *in vitro* system
- (for multi-color experiments) be sufficiently spectrally separated that they may be imaged individually without cross-excitation

This is a considerable list of necessary attributes, and there are some further desirable ones. For example, some fluorophores are more photodamaging than others, and some laser lines are also more damaging to cells and tissues than others. Fluorophores may be sensitive to pH or ionic strength and thus be inappropriate for the system of interest. Fluorescent proteins that are expressed *in vivo* are often describedas being either definitively "monomeric" or non-monomeric. Non-monomeric fluorophores will have more of a propensity to form homo-oligomers, and if imaging freely diffusing proteins may seed aggregation. In some cases, therefore, care must be taken to choose the monomeric form of the protein, the nomenclature for which is a lowercase "m" before the fluorescent protein name. For example, the monomeric form of the commonly used green fluorescent protein (GFP) is mGFP [30] that has an A206K mutation that suppresses putative dimerization between GFP molecules. Table 2 lists commonly used fluorophores alongside their usual applications.

| Fluorophore class | Example fluorophores | Applications | Notes |
|---|---|---|---|
| First generation fluorescent proteins | Green fluorescent protein [31], yellow fluorescent protein [32], red fluorescent protein [33] | *In vivo* protein labelling through genomic integration; FRET, *in vitro* labelleing, STORM | Not all suitable for single-molecule imaging e.g. cyan fluorescent protein. Derived from sea anemones (RFP) or jellyfish (GFP) |
| Second generation red fluorescent proteins (mFruits) | mCherry, mOrange, mStrawberry [34] | As above | Increased brightness over first generation RFPs |
| Second generation GFPs | Enhanced GFP [35], monomeric GFP [30], superfolder GFP [36] | As above | Improved brightness, reduced dimerization, and quickly-maturing, respectively |
| First generation cyanine dyes | Cy3 (orange), Cy5 (far-red) [37] | Labelling nucleic acids, proteins, both *in vitro* and *in vivo*; FRET | Can be chemically conjugated to proteins, not genomically integrated. Sensitive to local conditions |
| Alexa Fluor family | Alexa Fluor 488 [38] | As above | Second generation of xanthene, cyanine, and rhodamine dyes with improves brightness and photostability |
| Hoechst | Hoechst 33342 [39] | Minor groove binding DNA stain | Excited by UV light |
| Janelia Fluor family | JF525 [40] | Cell permeable dyes, used *in vivo* with protein labelling such as Halo Tag/Snap Tag | Improved quantum yields, *ca.* 2x brighter than comparable first-generation cyanines |
| Photoactivatable fluorescent proteins | PAGFP [41], PA-mKate2 [42] | PALM *in vitro* and *in vivo* | Enters fluorescent state on application of UV lightPALM *in vivo* |
| Photoconvertible fluorescent proteins | mEos2 [43], Kaede [44], Dendra 2 [45] | As above | Change color on application of UV light |

**Table 2: Table of commonly used classes of fluorophores and their principal applications.**

## 5. Ideal properties of a super-resolution microscope relevant to localization microscopy

In Section 4 we discussed the properties of a good fluorophore for super-resolution imaging. Here we will briefly describe the properties of a good super-resolution microscope appropriate for localization microscopy.

Lenses used should be clean and ideally coated with an anti-reflective coating for the wavelengths used, and care must be taken to ensure they are mounted truly perpendicular to the optical axis. Lasers should be able to produce a few milliwatts at a minimum and should produce stable output. Mirrors should be rated for the correct wavelength – what is reflective for infrared may be largely transparent to visible light. The whole system should be mounted on an air table to reduce mechanical vibration. Cameras should be capable of acquiring at the desired speeds, e.g. 10ms/frame, and should be cooled to reduce shot noise. It is necessary also that the camera have some gain function to amplify the light collected, for example electron multiplying (EM) gain which produces a cascade of electrons to hit the CCD and thus enhance the signal – but also the noise. In general, for best fitting of single molecule spots the camera should be imaging at a resolution of approximately 40-60nm/pixel. This is a key consideration and may necessitate additional optics prior to the camera to expand the imaged light.

The objective lens is one of the key components. For best performance this lens should have a high numerical aperture and be ideally oil-immersion (that is, oil is placed between the coverslip and the objective lens) to ensure good optical contact and enable high photon capture. That said, for imaging in excess of a few microns depth a water immersion lens may mitigate potential issues of spherical aberration that occur with oil immersion lenses, but with the caveat of a reduced numerical aperture of ~1.2, that reduces the photon capture budget. For dual-color experiments, chromatic aberration can be a problem – red and green light for example will come to focus at slightly different distances by a simple non-achromatic lens and therefore at a given height a red fluorophore may be in focus and a red fluorophore slightly defocused. There are four principal ways to get around this. i. One can measure the chromatic aberration and correct for it in image post-processing. ii. One can use an automatic stage with multiple settable heights and move between focal distances between acquisitions. iii. One may acquire all the green fluorescence data, manually refocus, and then take all the red fluorescence data (for example). iv. One may purchase an objective lens that is apochromatic and has minimal chromatic aberration. The first three techniques have drawbacks – careful calibration is needed for correcting chromatic aberration in post-processing, automatic stages may suffer from drift, and acquiring the data separately is non-trivial on unfixed samples since the acquisition may take some time. Moreover, the first acquisition may damage the system before you get the chance to look at the second fluorophore. Overall, if resources permit, an apochromatic lens is the best method for multi-color experiments. Though each optical microscope is different, the basic principles are similar across all, and a sample schematic is given in Figure 4. For convenience we will refer here only to an epifluorescence microscope, which is one in which the excitation light goes through the back of the objective and enters the sample collimated. The path the light takes is as follows: first, it is emitted from the laser, and if there are multiple lasers present then the fluorescence emissions are combined using dichroic mirrors to form one beam. This beam is then expanded using a telescope and then propagates through a lens that focusses the light on the back focal plane of the objective lens. The beam is directed into the objective lens by way of a dichroic mirror that reflects the excitation light but not the emitted light. The light hitting the objective lens then has originated from one focal distance away from the objective and is therefore collimated after the objective. This process of focusing and collimating with the objective is a second telescope that has the effect of reducing the beam width considerably. To get a beam of the correct width at the sample the expansion of the laser by the first telescope can be varied. After the excited fluorophores have emitted photons, a proportion of these is captured by the objective lens once more and collimated by it. The light path then goes back down the microscope and this time through the dichroic mirror and is focused typically on the back port of the microscope, where it may be re-collimated or imaged directly. For a dual-color experiment, the different color light must be split and imaged separately, either on two cameras or on separate parts of the same camera chip, or potentially using time-sharing just as with Alternating Laser Excitation [46] that uses a multi-bandpass filter set. In Figure 4b we show the simplest setup, a static color splitter based on a dichroic mirror imaging each channel on a separate part of the detector.

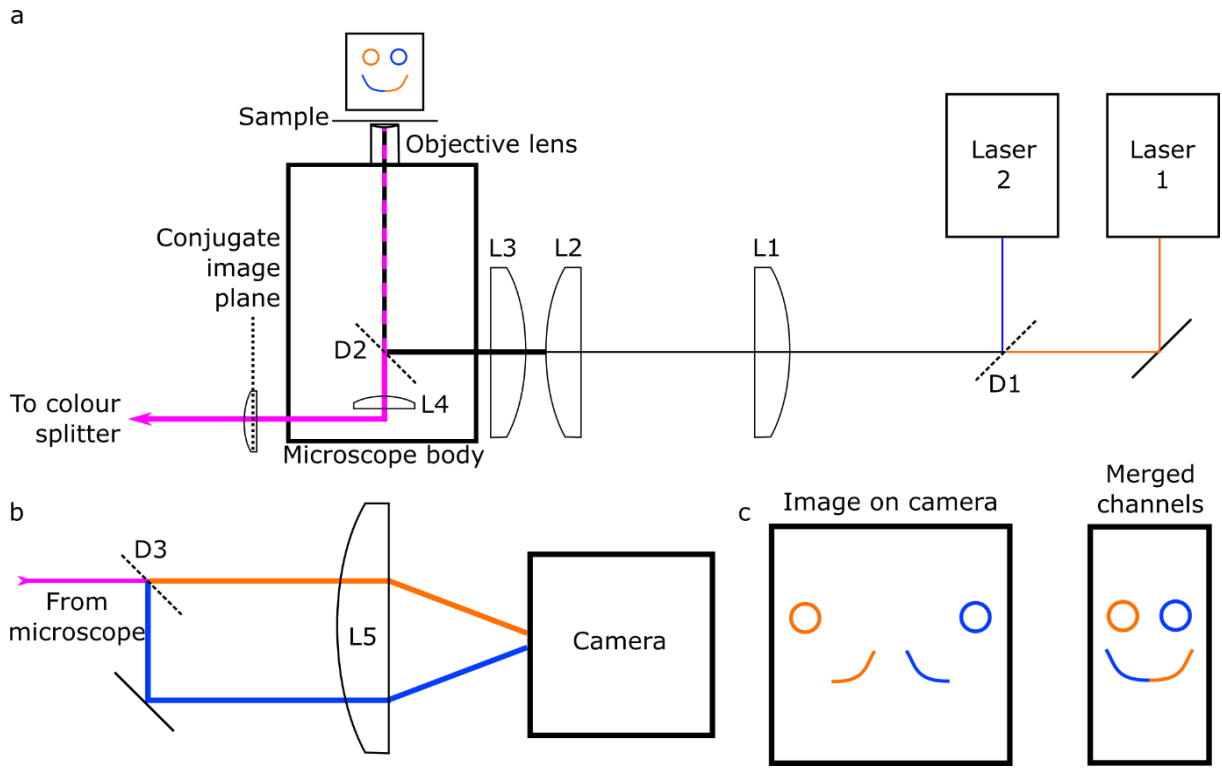

Figure 4. Schematic of a super-resolution STORM/PALM microscope. The lasers are combined by the dichroic mirror D1 and the beam is expanded by the lens pair L1 and L2. The lens L3 just before the microscope body forms a telescope with the objective lens and ensures the beam is the correct width and comes out of the objective collimated. Excitation light is directed into the objective using the dichroic mirror D2, which allows the captured fluorescence (pink) though. The imaged light is then focused on the side port of the microscope with the lens L4 within the microscope housing itself, though for convenience we do not image here but recollimated the imaged light with a lens placed at the conjugate image plane (marked). b) Principles of a color splitter. Collected light is passed through the dichroic mirror D3, which separates the two channels (here orange and blue). The distinct channels are focused on the camera chip with the lens L5, so that each color channel hits a separate half of the chip as seen in panel c. c) separate channels may be merged to recover the true image.

## 6. 3D localization

Localization in 3D can be approached in one of two ways. Firstly, a sample can be scanned through in the *z* direction building up a full 3D stack of the entire system of interest. Then the *z* position can be approximated as being the slice in which a given PSF is most in focus. This has the distinct disadvantage of being extremely slow – the best frame rates are around 2 full stacks per second, but almost all cellular processes happen three orders of magnitude quicker than that. For fixed samples this may be appropriate but for understanding dynamical processes it is simply inappropriate.

Instead, the PSF of the fluorophores can be altered through lenses or spatial light modulators (SLMs) so that they are non-symmetric about the focal point in *z*. Two principal techniques for this have emerged, namely astigmatism microscopy [47] and double-helix point spread function (DH-PSF) imaging [48].

For astigmatism imaging, the emitted light from the sample propagates through a cylindrical lens between the microscope and camera. This modifies the PSF from being rotationally symmetric – a Gaussian profile – into more of an elliptical profile. The orientation of the ellipse is dependent on whether the fluorophore is above or below the focal plane when imaged, and the ratio of the major to minor axes depends on the specific distance, as shown in Figure 5, which shows a simulated fluorophore's appearance as a function of *z* position. For this to work in practice, before experiments an *in vitro* fluorophore sample should ideally be imaged and scanned in *z* in known increments using an automated nanostage. The ratio of vertical to horizontal axis may then be measured for each slice and plotted against vertical distance. A fluorophore's focal point is where the ratio is 1, so that the relative absolute distance from the focus can be found. When imaging in an experiment, the focal plane is set and kept constant and the *z* positions of the fluorophores measured relative to that. In practice, this look-up table-based methodology is robust and requires only one additional lens in an existing fluorescence microscope, while the fluorophores themselves have only the same constraints as for 2D imaging. To date, astigmatism imaging with fluorophores *in vitro* and *in vivo* has shown an ability to beat the axial resolution limit by approximately a factor of 2-3, with axial spatial precisions of *ca.* 50 nm being common [49].

DH-PSF is a more complex technique requiring considerable different optics and a reconfiguration of the imaging path of the microscope. In a typical design, emitted light is collected and reflected off an SLM, while the light itself is imaged at an angle of 30° from the emitted light's optical axis. This produces a PSF that forms a double helix along the optical axis. When imaged in 2D, this appears as two separate dots, whose orientations are dependent on the *z* position of the fluorophore as seen in Figure 6, which simulates the appearance through *z* of a double-helical PSF. An angle may be generated by finding the angle between the vector linking the two fluorescent spots and the *x* axis. Then, as for astigmatism imaging, the angle-to-distance lookup table must be generated ideally from *in vitro* fluorophores before production data is acquired. The *xy* position of the fluorophore is taken to be the center of the two spots. This can be calculated by finding the centers of the spots themselves by fitting a 2D Gaussian to each, or by finding the centroid of the two-spot system with a specific centroid algorithm, through this latter technique is computationally more costly. An important drawback of DH-PSF imaging is that to generate the double helix the light is effectively split in two, and each spot has half the brightness of the full fluorophore at a given *z* position. If one is working in a low signal-to-noise regime, this reduction in brightness may make the spots indistinguishable from background. However, the *z* axis resolution is excellent – with sufficient photons detected the localization precision can be below 6 nm [50].

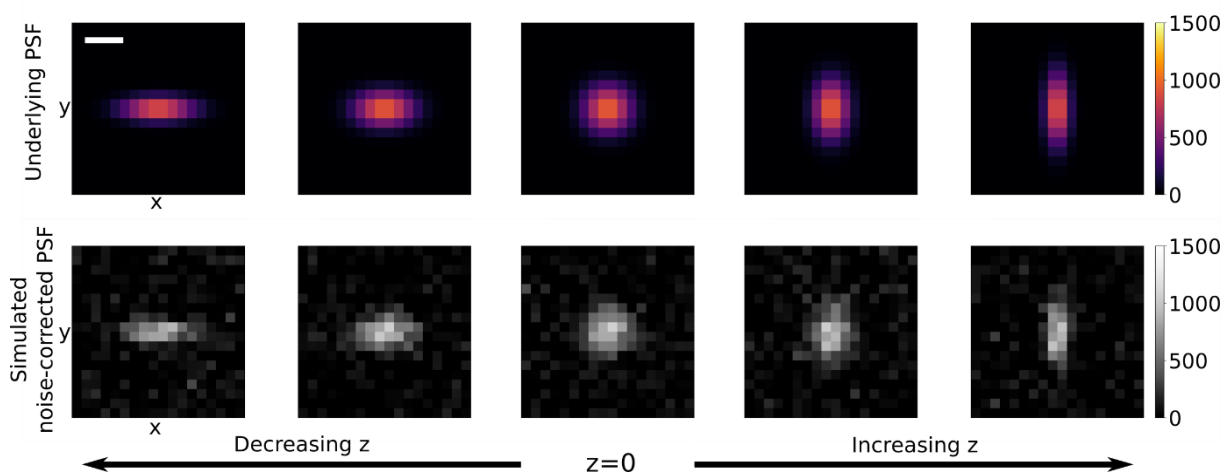

**Figure 5: astigmatism point spread functions with distance from the focal plane. PSF modelled as a 2D Gaussian with *x* and *y* sigma values set according to simulated "height". With each height step, σ$_x$ is reduced**

by 1 and $\sigma_y$ is increased by 1. For the lower panels, Gaussian noise has been added to each pixel to simulate background, but more complex noise such as camera shot noise are not included. Bar: 200 nm

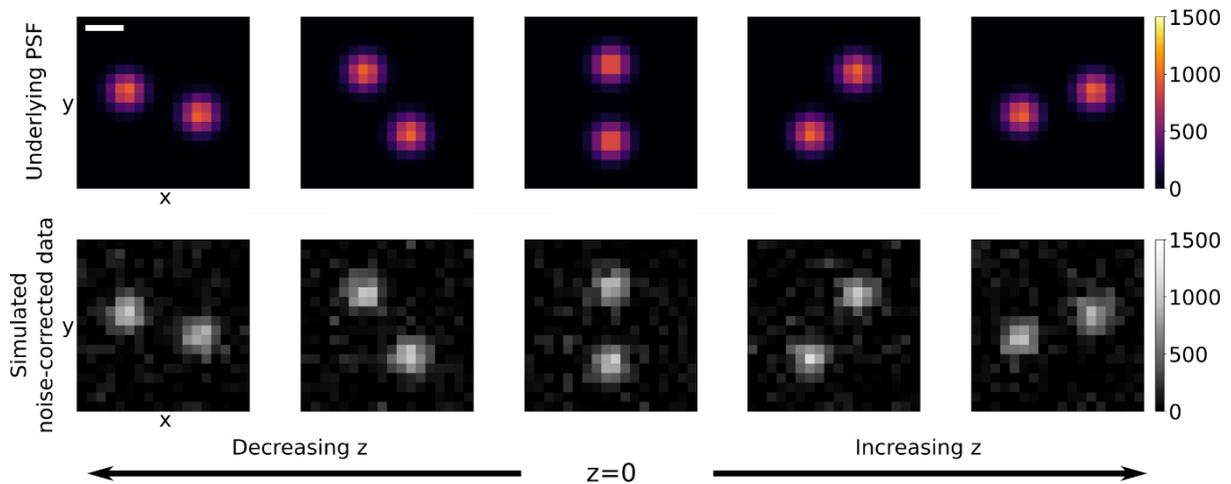

Figure 6: double helical point spread functions with distance from the focal plane. PSFs are simulated as two Gaussian distributions at opposite ends of an axis that rotates with each step in *z*. Again only Gaussian background noise is included in the lower panel simulations. Bar: 200 nm

## 7. Analyzing 2- and 3D localization data

Having found the trajectories of individual spots, a question immediately arises what to do with it. Broadly, we may define three categories of the trajectories that may each give useful information: position, velocity, and brightness.

Analysis of the positions themselves gives access to diffusion coefficients by comparison to Brownian motion, as well as colocalization information between molecules – i.e. tagging different targets with different color fluorophores, measuring the positions of each and identifying if they are in the same place. Simple positional analysis also may tell us if a protein is in the nucleus or cytoplasm, for example. As well as these, the overall spot distribution may be analyzed to determine if there are identifiable distinct regions to which multiple fluorophores belong. This suite of techniques is known as cluster analysis.

Various methods exist to perform cluster analysis. Most straightforward are distance-based methods such as the Voronoi method [51]. This generates a set of regions around each PSF such that each region is the area closer to the seed PSF than to any other, with small regions then indicating a cluster. Also widely used are density techniques such as density based spatial clustering of applications with noise (DBSCAN) [52] which iterates across all localizations and assesses the local density within a set radius *r*. If there is a minimum number of spots within the circle defined by *r* a cluster is accepted. This continues until the boundary of the dense area – and thus the entire cluster – is found. This repeats through all spots to find all clusters. Similarly using density are pure statistics-based methods of measuring clustering, particularly Ripley's H, K, and L functions [53]. These are a group of well-defined statistical transforms which can be applied to the image data and which have minima and maxima correlating to how clustered the data is. The values of the functions however only indicate whether over the spatial extent analyzed clustering is indicated. To classify points into discrete clusters requires analysis beyond the functions. This could be done by using the extended L function as proposed by Getis and Franklin [54], [55]

More recently, Bayesian analysis techniques have been developed which make use of advanced statistical models to evaluate clustering and in general aim to remove the level of human input or parameter selection needed during analysis. Bayesian implementations often make use of the statistical functions described above [56] and the number of clusters is then predicted with reference to the model, usually that the clusters are approximately spherical with molecules inside the cluster distributed according to a Gaussian [57]. Bayesian approaches are also valuable for determining the mode of molecular mobility of tracking data in localization microscopy, for example, are molecules freely diffusing or confined [18].

In principle, the outputs of these deterministic methods can also be used to train machine learning models. However, machine learning is often sensitive to the input – if the data to be analyzed is too dissimilar from the training data the output will be unreliable at best. One implementation of neural networks for cluster analysis was published in 2020, which used a neural network trained on a given number of nearest-neighbour distance values, showing efficient computational performance compared to Bayesian methods or DBSCAN on both simulated and experimental data [58]. However, the extensive training needed may offset this gain depending on the size of the dataset to be analyzed.

Velocities may also be characterised, and this is most commonly done in the context of molecular machines, where the step sizes and overall movement speed are difficult to determine by any other means and yet are crucial to biological function. By tracking fluorescently tagged molecular machines or cargoes these parameters can be accurately determined. Similarly, the overall drift of diffusing molecules may be examined to understand whether the Brownian motion they are undergoing is directed (for example facilitated diffusion, or an active process requiring the input of external free energy) or whether it is truly a random walk.

Finally, the intensity (i.e. brightness) of the fluorescent spots contains significant information about the system. Specifically, if we have a population of fluorescently-tagged molecules, we may analyze the distribution of intensities to uncover whether they are monomeric or aggregating into clusters. For systems where we know that aggregation happens – for example in liquid-liquid phase separation [59] – we can use the intensity through time to work out the total stoichiometry of molecules within the cluster. Whatever the purpose, the method for this is the same, and is done by taking the initial total intensity and dividing it by the intensity of a single fluorophore. The most important parameter to determine is thus the intensity of a single fluorophore which we denote as the Isingle value. By plotting the total intensity of a cluster through time we will see the decrease in intensity as the fluorophores in the cluster photobleach occurs in a step-wise fashion such that the size of a step, once noise is removed, is an integer multiple of the Isingle value. There may be differences of *ca.* a few tens of percent with estimates made *in vitro* for Isingle due to different in local excitation intensity, and buffering conditions inside a cell. Therefore, it is important to determine Isingle in the physiological context. The simplest way to achieve this is to use only the final photobleaching step where there is only one fluorophore that bleaches to leave only the background noise – more accurate than attempting to count all steps since these are limited to a maximum of 6-7 depending on the dye used. Further, steps involving more the one photobleached molecule in a sampling time window will have a higher associated noise due to Poisson sampling of photons at higher intensities. To obtain Isingle, the full intensity through time track can be fitted to a stepwise function usually such as a hidden Markov model [60], or other edge-preserving filters such as Chung-Kennedy [61], and the step sizes extracted and averaged – simulated data of an intensity track is shown in Figure 7a. Alternatively, the intensities of every spot can be plotted. If the was taken in the truly single molecule photoblinking regime, the majority of tracked spots should be single molecules, and therefore on a plot of intensity against number of spots a peak would be expected around Isingle. This overall process is known as step-wise photobleaching and is suitable for analysing either *in vivo* or *in vitro* data with the proviso that the Isingle values should ideally be determined separately for each sample. An illustration of simulated intensity-time data is in Figure.

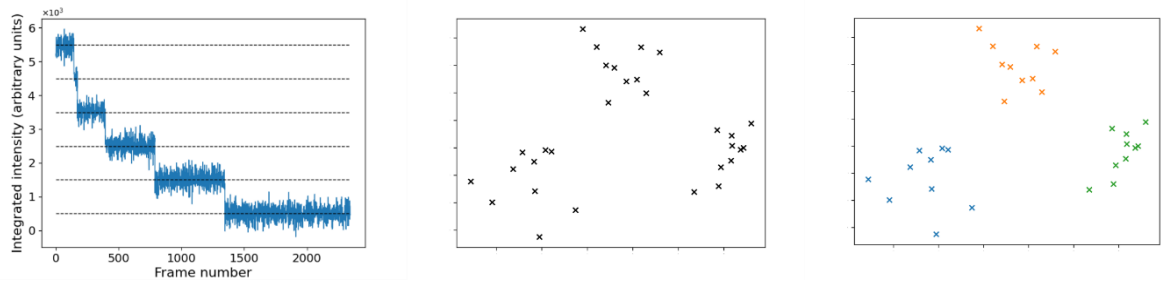

a. Figure 7 a) simulation of a step-wise photobleaching of a single intensity track; b and c) schematic of simple cluster analysis

## 8. Applications of localization microscopy

Single-molecule localization methods have been extensively applied both *in vivo* and *in vitro* to elucidate a wide range of biological processes. These include organization of molecules within the cell, the interplay between various cytoskeleton elements, and measuring diffusion coefficients. These details can tell us about what the key molecular interactions inside cells for specific biological process, as well as insights into mobility of molecular complexes and how these are influenced by the microenvironment of the cell. Here, we briefly present some biological results obtained to date.

DNA and RNA processes are amongst the most important in the cell and they have been studied extensively with single molecule tools. Yan *et al* used single molecule imaging to monitor mRNA translation and measure the switching between translating and non-translating states, finding translation repression due to specific sequences [62]. Also working on replication, Syeda *et al.* used dual-color imaging of the Rep helicase to demonstrate its dependence on PriC and the helicase's means of negotiating proteins bound to DNA [63]. Wooten *et al.* recently demonstrated super-resolution imaging could be used for epigenetic studies of chromatin fibers [64] in eukaryotes.

Away from DNA, localization microscopy has been used extensively to image the cellular cytoskeleton such as the organization of actin in 2D [65] and 3D [66] as well as the distribution and degradation of intermediate filaments [67] and intracellular trafficking dynamics where microtubules intersect [68]. Live cell imaging with 3D localization has shown colocalization of proteins and the cell surface [69], as well as the distribution of eukaryote transcription factors which may be used to map the overall genome [49]. 2D localization of synthetic sensors inside living cells has also recently been demonstrated which suggests that localization of proteins may soon be correlated with the physical conditions around them [70], while diffusion coefficient analysis has shown that under osmotic stress eukaryotes experience slower diffusive behavior in the cytosol [71]. Super-resolution microscopy has also been used to observe clustering of key eukaryotic proteins [72], and more generally are being used to understand the currently murky world of liquid-liquid phase separation [59].

Alongside this *in vivo* work, considerable progress has been made through *in vitro* experiments also. Protein aggregation can readily be studied in a microscope slide, and amyloid proteins implicated in Alzheimer's disease have been imaged aggregating in human cerebrospinal fluid [73]. Step-wise photobleaching has been used to understand aggregation of amyloid-β [60] *in vitro* also. DNA origami has been extensively studied for some time, and as well as imaging the structure of the origami tile, it has been used to more robustly characterize protein copy number using immunofluorescence [74].

## 9. Conclusion and future perspectives

Overall, it is clear that localization microscopy is an enormously valuable technique, enabling new insight into a range of complex biological process. Both PALM and STORM methods can be used for reconstructing the fine structure of biological structures, while fitting to diffusing molecules exposes diffusion coefficients, and subcellular organization in response to stress or during key biological processes. The detail we are now able to obtain using the methods in this chapter is immense. However, there remain technical challenges that need to be addressed, and there remain drawbacks of STORM/PALM-type experiments, for example in certain instances cells may need to be fixed, losing valuable dynamic information. As we look to the future of biological microscopy, the focus will increasingly be on multi-method and correlative approaches, which promise to give information beyond what is currently possible. Life does not exist separately to physics; rather, cells leverage physical laws to organize and regulate their internal conditions. With cutting-edge physical sensors now available to measure crowding, pH, and ionic strength, we may now begin to correlate our precise local data with the prevailing physical processes and conditions. This integrative understanding across disciplines will be the key battleground in the quest to understand life and develop the medical therapies of the future.

## 10. Take home message
- A fluorophore may be localized by fitting a PSF to an acquired image
- The best spatial precision in fixed cells is ~1 nm and in living cells ~30 nm for millisecond to tens of milliseconds time resolution
- Multiple molecules of interest may be labeled and imaged in multiple colors in the same cell that can enable insight into dynamic molecular interactions
- Photoblinking and localization of single molecules requires high laser power and so there is a trade-off with photodamage of samples
- Step-wise photobleaching can be used to determine molecule copy numbers and molecular complex/assembly stoichiometries
- Tracking molecular complexes can yield valuable information about the molecular mobility and the local microenvironment of cells
- Reshaped PSFs can enable 3D spatial information

## 11. Acknowledgements

This work was supported by the Leverhulme Trust (RPG-2019-156) and EPSRC (EP/T002166/1).